\documentclass[11pt,twoside]{article}

\usepackage{asp2006}
\usepackage{epsf}

\begin{document}
\title{Atomic Fluorescence and Prospects for Observing Magnetic Geometry
 Using Atomic Magnetic Realignment}
\author{Kenneth Nordsieck}
\affil{Department of Astronomy, University of Wisconsin - Madison, 475 N Charter St, Madison, WI 53706}

\begin{abstract}
Yan and Lazarian have proposed a new technique through which the magnetic field geometry in the diffuse interstellar medium, or in circumstellar matter, could be determined from the linear polarization of interstellar absorption or fluorescence emission lines from ions pumped by an anisotropic illuminating flux.  New long-slit spectroscopic observations of the reflection nebula NGC2023, obtained with the Southern African Large Telescope Robert Stobie Spectrograph (RSS), have detected a number of atomic fluorescence  lines of OI, NI, SiII, and FeII for the first time in a neutral medium.  A model which predicts these lines and others illustrates which lines would be appropriate targets for an RSS spectropolarimetric investigation of this new diagnostic.
\end{abstract}


%%% MAIN BODY OF TEXT GOES HERE. CONSULT "INSTRUCTIONS FOR AUTHORS USING
%%% LATEX2E MARKUP", SECTIONS 2.3-2.6 FOR HELP WITH EQUATIONS, FIGURES,
%%% AND TABLES.

\section{Magnetic Realignment: How to Observe It}
Atomic resonance/ fluorescence lines may be linearly polarized when an\-iso\-tro\-pic optical pumping produces an an\-iso\-tro\-pic angular momentum distribution (``alignment'') within the ground state \citep{yl2006,yl2007,yl2008} (hereafter YL06,YL07,YL08).  This occurs if the pumping photon rate is greater than the collision rate, for instance within 1-10 pc of an OB star.  In the presence of a magnetic field, ``magnetic realignment'' then occurs when the Larmor frequency is greater than the photon rate. For the ``saturated'' case (valid in the diffuse ISM, where any field $> 0.1$  $\mu$Gauss causes realignment), the polarization depends on the 3D geometry of the magnetic field, the ion ground state configuration and sometimes on the pumping spectrum.  In circumstellar matter, $10-10^4$ $\mu$Gauss is required for realignment, and the polarization can depend also on the field strength. Related effects have been seen in the sun \citep{s1994}, but have not yet been demonstrated elsewhere.  As a diagnostic of the magnetic field, magnetic realignment is potentially more powerful than the 21 cm Zeeman Effect, since it is sensitive to weaker fields and works in hot gas, and more powerful than dust alignment, since it is sensitive to 3D geometry, gas properties and velocity.  This paper explores observational aspects of this effect.

\begin{figure}[!ht]
\plotfiddle{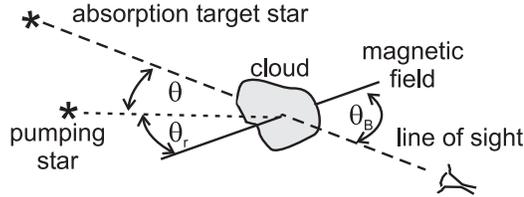}{0.75in}{0}{45}{45}{-140}{-250}
\caption{Geometry for Magnetic Realignment}
\end{figure}

Figure 1 shows the observational geometry.  A gas cloud with a magnetic field at an angle  $\theta_B$ from the line of sight is illuminated by a nearby star (the ``pumping star'') at an angle  $\theta_r$ from the field ($\theta$ from the line of sight).  In the absorption case, flux from a star behind the cloud (the ``target star'', which may be the same as the pumping star) is absorbed by ions in the cloud.  If the ions are aligned, the resultant absorption line is linearly polarized, with a position angle either parallel or perpendicular to the projected magnetic field, and a degree that depends on  $\theta_B$ and $\theta_r$.  For absorption, YL06 show that a non-zero polarization requires ions with at least three fine states in the ground level $(J \geq 1)$.  In neutral gas, the most common ions are NI, OI, SII, and FeII.  The spectral resolution required is
$R = \lambda / \Delta\lambda \ga 20,000$, to resolve the interstellar lines.  Since the absorption lines are almost entirely in the FUV, this requires a high resolution FUV spectropolarimeter \citep{n2003}.

In the emission case, the observed light is light from the pumping star which is scattered into the line of sight.  The unique polarization signature of magnetic realignment is a distortion of position angles of the scattered emission lines from the centrosymmetric pattern expected for pure reflection polarization. YL07 and YL08 show that ions need at least three fine \emph{or hyperfine} states in the ground level ($F \geq 1$) for a nonzero magnetic field effect.  This adds the resonance lines of NaI, KI, and AlII to the roster.  Also, transitions that do not return to the ground state (fluorescence) exhibit the effect, thus opening up numerous targets in the visible and NIR.  The spectral resolution required is more modest: for sensitivity against scattered dust continuum $R \ga 5,000$ is adequate.   Thus in principle, this effect can be observed with a ground-based moderate-resolution spectropolarimeter.  This paper describes the beginnings of such an investigation, targeted at fluorescence lines in the visible.

\section{Pilot Spectroscopic Observation: Fluorescence in NGC2023}
Visible-wavelength fluorescence of OI, NI, and SiII has been observed in HII regions and Planetary Nebulae \citep{g1976}.  However, these lines are faint compared to the recombination and collision-excited lines in these regions, and the existence of alternate excitation mechanisms confuses their use.  A better environment for the exploration of these lines is the neutral medium, for instance the gas within Reflection Nebulae, where photoexcitation is the sole excitation mechanism.  A rough calculation suggests that the lines should be visible above the dust-scattered continuum with an equivalent width of $0.1 - 10 \AA$, so should be visible with moderate resolution. No such observations have been published: we here report a pilot spectroscopic investigation of the bright RN NGC2023.

\begin{figure}
\plotfiddle{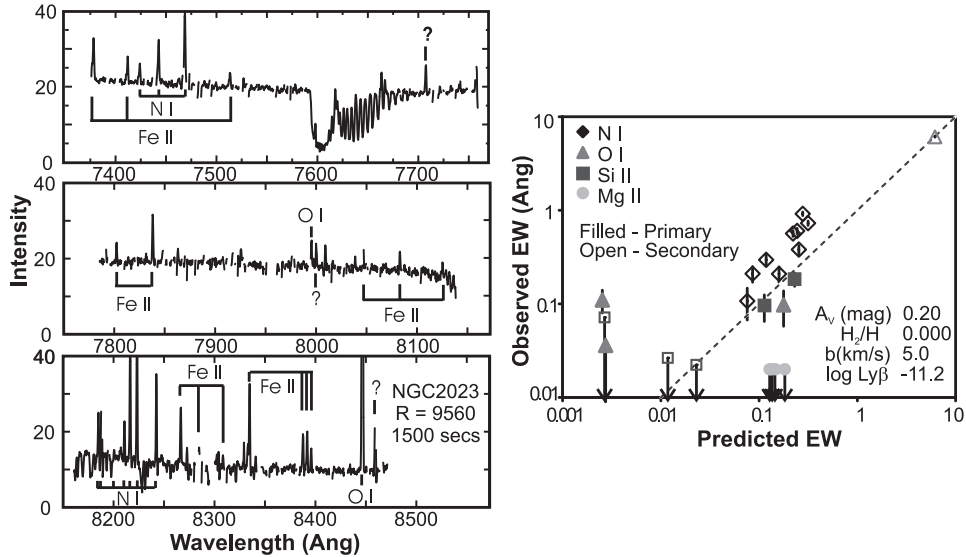}{2.7in}{0}{53}{53}{-210}{-55}
\caption{2a left: NGC2023 NIR spectrum, summed along the slit; 2b right: Observed vs predicted Equivalent Widths}
\end{figure}

Observations were obtained on 25-27 Oct, 2006 as a part of ``Performance Verification'' on the new Robert Stobie Spectrograph \citep{k2003} on the Southern African Large Telescope, a new 10m-class telescope (Buckley, Swart, \& Meiring 2006).  A North-South $0.6\arcsec \times 8\arcmin$ slit was offset $10\arcsec$ East of the central star, HD37903 (B1.5V).  Three different grating configurations were used, covering 3800-4560 \AA~($R = 7600$), 5425-6345 \AA~($R = 8500$), and 7375-8490 \AA~($R = 9560$) in first order.  Overall, 38 significant lines are seen: 7 unidentified, 4 primary fluorescence (the first photon emitted after excitation, the most useful for polarimetric purposes), 10 secondary, and 17 unclassified (FeII, and possibly CrII and TiII). Table 1 lists observed equivalent widths of the primary fluorescence lines, summed along the slit.  Figure 2a shows the near IR observation.  The line signal goes to zero $1\arcmin$ from the illuminating star, almost entirely interior to the well studied $H_2$ emission in the PhotoDissociation Region \citep{g1987}.  These lines provide a new velocity, temperature, and magnetic field probe of this warm neutral material.

\begin{table}[!ht]
\caption{Primary Fluorescence Lines}
\smallskip
\begin{center}
{\small
\begin{tabular}{cccccccccc}
\tableline
\noalign{\smallskip}
Ion & $\lambda$ (Air) & EW & pr/ob & Grnd & Upper & Final & $E_1$ & $U_\bot (max)$\\
& $\AA$ & $\AA$ & & & & & $\%$ & $\%$ \\
\noalign{\smallskip}
\tableline
\noalign{\smallskip}
SiII & 5957.56 & 0.095 & 1.19 & $^2P\onehalf$ & $^2S\onehalf$ & $^2P\onehalf$ & 50 & 0\\
SiII & 5978.93 & 0.184 & 1.25 & $^2P\onehalf$ & $^2S\onehalf$ & $^2P$\hbox{$\,^3\!/_2$} & 0 & 0\\
OI & 6046.44 & 0.110 & 0.02 & $^3P2$ & $^3S1$ & $^3P0,1,2$ & 0.00 & 0.00 \\
OI & 7995.07 & 0.098 & 1.82 & $^3P2$ & $^3D3$ & $^3P2$ & 24 & 7.57 \\
MgII & 8115.23 & $<0.020$ & 9.15 & $^2S\onehalf$ & $^2P$\hbox{$\,^3\!/_2$} & $^2D$\hbox{$\,^5\!/_2$} & 10 & 0\\
MgII & 9218.25 & (73.67) & (pred) & $^2S\onehalf$ & $^2P$\hbox{$\,^3\!/_2$} & $^2S\onehalf$ & 50 & 0\\
AlII & 8640.70 & (0.099) & (pred) & $^1S0$ & $^1P1$ & $^1S0$ & 22.2 & 1.04 \\
\noalign{\smallskip}
\tableline
\end{tabular}
}
\end{center}
\end{table}

A simple model has been constructed to verify that the observed lines are seen at approximately the expected strengths, and to predict other lines that might be seen in other parts of the spectrum.  There are 11 candidate ions for fluorescence, if we require a neutral medium (principle ion with IP $< 13.6$ eV), moderate abundance (Abundance / H $> 10^{-10}$), pumpable (first resonance $ < 13.6$ eV), and permitted visible-wavelength emission: NI, OI, MgII, AlII, SiII, ArI, TiII, CrII, MnII, FeII, and NiII.  We observe NI, OI, SiII, FeII, and possibly TiII and CrII.  The observed equivalent widths are adequately predicted by a simple shell illuminated by a B1.5V star, $A_V \sim 0.2$ mag, $b \sim 5$ km/s, and no H$_2$ absorption: the ratio of the predicted to observed equivalent widths are given in Table 1 and shown in Figure 2b.  All the predicted lines for our wavelength coverage are seen to within a factor of two of prediction, except for OI $\lambda$6048, which may in fact be FeII, and for the MgII lines $\lambda\lambda$8115/8120, which are not seen, perhaps due to additional depletion.  Outside of our wavelength coverage, we predict a detectable line of AlII at 8641 \AA~and two very strong lines of MgII at 9221 and 9244 $\AA$ (Table 1).  We have not yet made predictions for TiII, CrII, and FeII, because their non-LS coupling results in far more complex pumping.

\section{Expected Fluorescence Polarization Signals}
We can now predict polarization signals from the lines seen and predicted for NGC2023.  The polarization is the sum of the reflection polarization - the polarization expected from an unaligned (``thermalized'') ground state - plus that due to ground state alignment.  \citet{s1994} gives the reflection polarization,
\begin{equation}
p(\theta) = \threequarters E_1 \sin^2 \theta  /
(1 - \onequarter E_1 + \threequarters E_1 \cos^2 \theta )
\end{equation}
where $\theta$ is the scattering angle and $E_1$ is the ``polarizability'', which depends only on the angular momentum $J$ (or $F$) of the ground, upper and final levels (Table 1).  The polarization position angle is centrosymmetric: either perpendicular or parallel to the scattering plane. The maximum polarization is $\sim$$E_1$ at $\theta = 90\deg$ ; $E_1 = 1$ corresponds to Thompson scattering.

\begin{figure}[!ht]
\plotfiddle{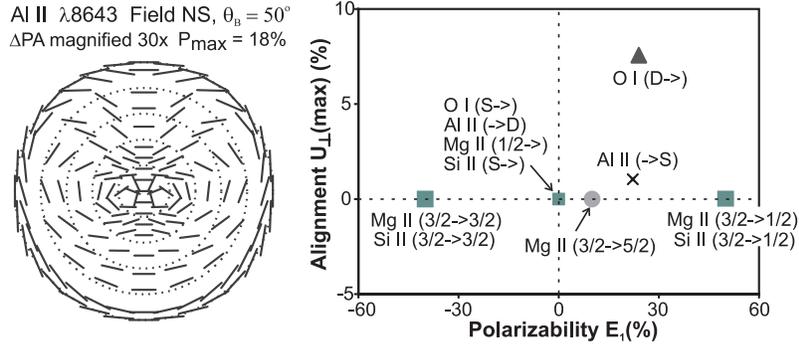}{1.6in}{0}{40}{40}{-160}{-94}
\caption{3a left: Polarization map for a spherical shell nebula, AlII $\lambda$8643; 3b right: Alignment polarization \textit{vs} reflection polarization for NGC2023 lines}
\end{figure}

The alignment polarization will be nonzero only for alignable ions, where the ground state is aligned by pumping via all the possible UV resonance transitions.  Addition of the alignment to the reflection polarization changes both the degree of polarization \emph{and the position angle}, depending on the scattering angle and the 3D magnetic field orientation. This results in a position angle that is no longer centrosymmetric, a feature unique to this process (see Figure 3a).  A useful measure of this signal is the deviation $\Delta PA$ from centrosymmetric, expressed as a polarization,
$U_\perp (\theta ,\theta_B) = p \sin 2 \Delta PA$.

A plot of~$\,U_\perp$(max) \textit{vs} $E_1$ (Table 1; Figure 3b) for each line is useful to plan the analysis. Of the ions we see or predict, two are not alignable (MgII and SiII), and 5 are (OI, AlII, TiII, CrII, and FeII). To deduce the magnetic field geometry at any position in a nebula, we need first to estimate the scattering angle $\theta$. For this we use lines of nonalignable ions with large $E_1$, like Mg II $\lambda$9218 ($E_1 = 50\%!$) with equation (1) to derive $\theta$.  Lines with zero polarizability (MgII $\lambda$9244, Si II $\lambda$5979) are used to measure and subtract foreground interstellar polarization.  Finally, to derive the magnetic field map, we use the map of $U_\perp$ for lines of alignable ions like OI $\lambda$7995 and AlII $\lambda$8641. (If there is significant variation in $\theta$ along the line of sight, a more complex model will be required).

A important consideration is the role of optical depth of the UV pumping lines.  If the pumping line is ``trapped'' (return is much more probable to the ground state than to an excited state), then when  $\tau_{UV} \gg 1$ the excitation will be dominated by diffuse trapped light, which is no longer anisotropic; any fluorescence from this level will be unpolarized, and the ground state cannot become aligned.  On one hand, most fluorescent exciters are never trapped, because the fluorescent transition itself has sufficiently high probability that the exciting radiation is rapidly destroyed: the maximum number of scatterings $\sim \tau_{max} \sim 1/$(escape prob) \citep{h1964}.  (In the simple model for NGC2023 above, the fluorescence intensity is modeled as being proportional to $\tau_{UV}$ or $\tau_{max}$, whichever is smaller). Thus the reflection polarization of fluorescent lines should not usually be depolarized by a large factor.  On the other hand, many alignable ions have lower lying trapped resonance lines that dominate the pumping of the ground state, whose effect is then to depolarize the \emph{alignment} polarization.  Remedies for this situation are to look at thinner nebulae, and to look at ions which do not have trapped resonance lines, such as FeII.

\section{Future Work}
We plan spectroscopy of NGC2023 and other nebulae to look for the predicted MgII and AlII lines, followed by slit spectropolarimetry to find the best lines.  The new Fabry-Perot polarimetry mode of the SALT RSS spectropolarimeter will then be used for construction of a magnetic field map.  Modeling improvements will include an FeII realignment calculation and fluorescence model.

%\subsection{}   %%% Second level section head (remove "%" symbol)
%\subsubsection{}   %%% Lowest level section head (remove "%" symbol)
%\section*{}    %%% Unnumbered top level section head (remove "%" symbol)
%\subsection*{}   %%% Unnumbered second level section head (remove "%" symbol)
%\acknowledgements %%% Text of acknowledgements runs on after this command.

%%% THE BIBLIOGRAPHY
%%%
%%% CONSULT SECTION 3 OF "INSTRUCTIONS FOR AUTHORS" FOR HOW TO USE NATBIB.
%%% AUTHORS ARE ENCOURAGED TO USE EITHER THE "THEBIBLIOGRAPY" ENVIRONMENT
%%% BY UNCOMMENTING (DELETING THE "%" SYMBOL) THE COMMANDS BELOW, OR BY
%%% USING THE BIBTEX ENVIRONMENT. TO FIND OUT WHICH IS APPLICABLE TO YOUR
%%% CONTRIBUTION, CONSULT THE VOLUME EDITORS FOR YOUR PROCEEDINGS.
%%%


\begin{thebibliography}{}
\bibitem[Buckley, Swart, \& Meiring(2006)]{b2006}
Buckley D.A.H., Swart G.P., Meiring J.G., 2006, Proc SPIE, 6267
\bibitem[Gatley et al.(1987)]{g1987}
Gatley, I., Hasegawa, T., Suzuki, H., Garden, R., Brand, P., Lightfoot, J., Glencross, W., Okuda, H., \& Nagata, T. 1987 \apj, 318, 73
\bibitem[Grandi(1976)]{g1976}
Grandi, S.A. 1976 \apj, 206, 658
\bibitem[Hummer(1964)]{h1964}
Hummer, D.G. 1964 \apj, 140, 276
\bibitem[Kobulnicky et al.(2003)]{k2003}
Kobulnicky, H.A., Nordsieck, K.H., Burgh, E.B., Smith, M.P., Percival, J.W., Williams, T.B.and O'Donoghue, D. 2003, Proc SPIE, 4841, 1634
\bibitem[Nordsieck et al.(2003)]{n2003}
Nordsieck, K.H., Jaehnig, K.P., Burgh, E.B., Kobulnicky, H.A., Percival, J.W., \& Smith, M.P. 2003, Proc SPIE, 4843, 170
\bibitem[Stenflo(1994)]{s1994}
Stenflo, J. O. 1994, Solar Magnetic Fields (Dordrecht: Kluwer)
\bibitem[Yan \& Lazarian(2006)]{yl2006}
Yan, H., \& Lazarian, A. 2006, \apj, 653, 1292
\bibitem[Yan \& Lazarian(2007)]{yl2007}
Yan, H., \& Lazarian, A. 2007, \apj, 657, 618
\bibitem[Yan \& Lazarian(2008)]{yl2008}
Yan, H., \& Lazarian, A. 2008, \apj, 677, 1401
%\bibitem[]{}
%\bibitem[]{}
\end{thebibliography}
\end{document}